\documentclass[11pt]{iopart}

\usepackage[bookmarks]{hyperref}
\usepackage[dvips]{graphicx}
\usepackage{graphics}
\usepackage{exscale}
\usepackage{epsfig}
\usepackage[T1]{fontenc}
\usepackage{bm}
\usepackage{bbm}

\usepackage{version}

\usepackage{latexsym}

\renewcommand{\epsilon}{\varepsilon}

\newcommand{\expval}[2]{ \langle  #1 #2\ \!\! \rangle}

\newcommand{\elcre}[2]{ c^{\dagger}_{#1,#2}}
\newcommand{\elann}[2]{ c_{#1,#2}}

\newcommand{\vk}{{\bm k}}

\newcommand{\hc}{\mathrm{h.c.}}

\includeversion{commentst1} 

\begin{document}
\title[Probability distribution of fluctuations in
  Anderson-Holstein and Holstein-Hubbard Models]{Numerical Renormalization
  Group Study of Probability Distributions for  Local  Fluctuations in the
  Anderson-Holstein and Holstein-Hubbard Models} 
\author{Alex C. Hewson${}^1$ and Johannes Bauer${}^2$}
\address{${}^1$Department of Mathematics, Imperial College, London SW7 2AZ,
  United Kingdom}
\address{${}^2$Max-Planck Institute for Solid State Research, Heisenbergstr.1,
  70569 Stuttgart, Germany} 




\begin{abstract}
We show that information on the probability density of
local fluctuations can be  obtained from a numerical renormalisation group
calculation of a reduced density matrix.  We apply this
approach to the Anderson-Holstein impurity model to  calculate the ground
state probability density $\rho(x)$ for the displacement $x$ of the local
oscillator. From this density we can deduce an effective local potential for
the oscillator and compare its form with that obtained from a semiclassical
approximation as a function of the coupling strength.
The method is extended to infinite dimensional Holstein-Hubbard model using
dynamical mean field theory. We use this approach to compare the
probability densities for the displacement of the local oscillator in the
normal, antiferromagnetic and charge ordered  phases.  
\end{abstract}
\maketitle

\section{Introduction}

The numerical renormalization group \cite{Wil75,BCP08} (NRG) approach has been successfully
applied to the calculation of static and dynamical response functions for 
 models of magnetic impurities and quantum dots, and
also been applied to lattice models, such as the Hubbard model, in the framework of dynamical mean field
theory (DMFT) \cite{GKKR96}. The response functions calculated give information about the
low energy fluctuations which are of particular interest in the strong
correlation regime. However,  as a thermodynamic average is taken in
calculating these response functions, some of the information about these
fluctuations, which is contained in the original many-body states, is
lost. Here we show that if we take only a partial thermodynamic average, such
as in the calculation of a reduced density matrix, we can learn more about the
nature of the local fluctuations, and how they vary as a function of the
interaction terms.\par 

We illustrate the approach first of all for the Anderson-Holstein impurity
model. In the Anderson-Holstein model the occupation of the impurity state is
linearly coupled to a local harmonic oscillator, which has spatial coordinate
$x$, representing the lattice degrees of freedom around the impurity site. As
the coupling of the oscillator to the impurity state is increased, the nature
of the local fluctuations of the oscillator changes. If we take a full
thermodynamic average then we only get averaged information about the
oscillator. If we take a partial thermodynamic average, and calculate the
reduced density matrix at the impurity site, treating the rest of the system
as an 'environment',  we can learn more about how the local lattice
fluctuations vary as the coupling strength increases. 
From the reduced density matrix, calculated using the NRG, 
we can deduce the  probability distribution $\rho(x)$ for the  $x$
coordinate of the oscillator. From  $\rho(x)$ we can also deduce an effective
potential $V_{\rm eff}(x)$, and study the change of this potential as a
function of the interaction strength and of the frequency of the local
oscillator. Later in the paper, we extend the method to the infinite
dimensional Holstein-Hubbard lattice model using dynamical mean field theory
(DMFT) in combination with the NRG. We then compare how the local  probability
distribution $\rho(x)$ changes in the normal, antiferromagnetic  and charge
ordered states for this model.

\section{Local Fluctuations in the  Anderson-Holstein Impurity Model}
\label{sec:anders-holst-impur}

 The  Anderson-Holstein  model corresponds to the single impurity Anderson
 model~\cite{And61} with an
additional 
linear coupling to a local phonon mode, as in  the Holstein
model~\cite{Hol59}. The Hamiltonian takes the form,  
\begin{eqnarray}
&&H=\sum_{\sigma}\epsilon_f\hat n_{f,\sigma}
    +U\hat n_{f, \uparrow}\hat n_{f, \downarrow} + g (b^{\dagger}+b)
   (\sum_\sigma {\hat n_{f,\sigma}} -1) \nonumber \\
&&+\sum_{{\vk},\sigma}V_{\vk}(c^{\dagger}_{f,\sigma} c^{}_{{\vk}\sigma}
    + \hc)
+\sum_{{\vk}\sigma}
    \epsilon_{\vk}c^{\dagger}_{{\vk}\sigma}c^{}_{{\vk}\sigma}
    + \omega_0 b^{\dagger}b.  
\label{ah}
\end{eqnarray}
The impurity  level $\epsilon_f$, as in the usual Anderson model, is hybridised with
conduction electrons of the host metal via a matrix element $V_{\vk}$, with
an interaction term $U$ between the electrons in the localized f (or d)
state.  There is in addition a coupling $g$ of the impurity site  occupation
$\hat n_{f,\sigma}=c^{\dagger}_{f,\sigma} c^{}_{f,\sigma}$ 
to a local
oscillator of frequency $\omega_0$. A measure of the hybridisation is the
width factor, $\Delta(\omega)=\pi\sum_{\vk}V_\vk^2\delta(\omega-\epsilon_{\vk})$,
which for a flat conduction band of width $W=2D$, and $V_\vk$ independent
of $\vk$, we can take as a constant $\Delta=\pi V^2/2D$.  The oscillator coordinate operator $\hat
x$ in terms of the creation and annihilation operators, $b^{\dagger}$ and $b$, is given by
$\hat x=(b+b^{\dagger})/\sqrt{2\omega_0}$, where we have taken the mass of the
oscillator, $M=1$. We have also set $\hbar=1$ so that $x^{-1}$ has the dimension of
of square root of energy, and it is convenient to define characteristic length
scale by $x_0=1/\sqrt{\omega_0}$. A convenient measure of the effects of the
electron-phonon coupling on the electronic system is the parameter
$\lambda=2g^2/\omega_0$. 
In the limit $\omega_0\to\infty$, such that $\lambda$ remains finite, the model maps
into the Anderson model with $U\to U-\lambda$, $\epsilon_f\to \epsilon_f
+\lambda/2$.
The behaviour of the model has been studied using the NRG \cite{HM02}
and it has been used to study the transport through a quantum dot in the
presence of a coupling of the occupation dot to local phonon modes \cite{CNG04}.

To learn more about the state of the oscillator we use the NRG to calculate
the reduced density matrix $\rho_{\rm red}$ at the impurity site. A procedure
for calculating the reduced density matrix was introduced into NRG
calculations by Hofstetter \cite{Hof00}. The original motivation was to find an improved way of
calculating
the higher energy features in the spectral density of the impurity Green's
function in cases of broken symmetry. For NRG calculations the system is
recast in the form of a linear chain with impurity at one end. Sequential
diagonalisations are then carried out starting at the impurity site.
The information from the shorter length chains is used
to calculate the higher energy features in the spectral density,
and the longer chains the low energy features. In the case of broken symmetry
the ground state for the shorter chains underestimates the degree of symmetry
breaking. In Hofstetter's modified procedure the ground state is first
estimated
from the longest chain calculated, and then used to deduce the density matrix of 
the sites corresponding to shorter chain lengths.  Incorporating the reduced
density matrix into the calculation of the spectral density from the shorter
chains corrects the deficiencies of the standard approach in the case of
broken symmetry. Refinements of this approach have been introduced more
recently based on the use of a complete set of NRG states
\cite{PPA06,WD07}. These have the advantage that the sum rules on the total
spectral density are satisfied exactly, rather than approximately as in
earlier versions of the NRG approach. \par   
 
The calculation of the density matrix gives additional
information which we can exploit. For example in the Anderson-Holstein model,
if we work backwards from the longest chain, we can deduce
the reduced density matrix at the impurity site $\rho_{f,{\rm red}}$. The
matrix elements of this reduced density matrix will be with respect to the
basis states at the impurity site, the states of all the other sites are
averaged over as they are taken to be part of the environment. The electronic
states at the impurity site can be labeled by the local charge $q$
($q=\sum_\sigma n_{f,\sigma}$), and the $z$-component of spin $m_z$, and the
index of the harmonic oscillator states $\nu$. The 
matrix will be diagonal  with respect to the spin and charge indices, and so 
a typical matrix element can be expressed as  $(\rho_{f,{\rm
    red}}(q,m_z))_{\nu,\nu'}$. 
The probability  distribution function $\rho(x)$ for the oscillator coordinate
$x$ is given by 
\begin{equation}
\rho(x)=\sum_{q,m_z} \rho(x:q,m_z)
\label{rho}
\end{equation}
where
\begin{equation}
\rho(x:q,m_z)=\sum_{\nu,\nu'} \phi_{\nu}(x) (\rho_{f,{\rm
    red}}(q,m_z))_{\nu,\nu'}\phi^*_{\nu'}(x),
\label{rho2}
\end{equation}
and $\phi_{\nu}(x)$ is the normalised real space harmonic oscillator wavefunction,
\begin{equation} 
\phi_{\nu}(x)
  =\left({\sqrt{\omega_0}\over\sqrt{\pi}2^\nu
  \nu!}\right)^{1/2}e^{-\omega_0x^2/2}H_\nu(\sqrt{\omega_0}x),
\end{equation}
with the Hermite polynomial $H_\nu(x)$  of order $\nu$.

It is possible to define an effective potential for the oscillator using
an effective wavefunction defined by $\psi_{\rm eff}(x)=\sqrt{\rho(x)}$,
which is taken to be a solution of the one-dimensional Schr\"odinger equation
(recall $M=1$, $\hbar=1$),
\begin{equation}
-\frac{1}{2}\frac{d^2\psi_{\rm eff}(x)}{dx^2}+V_{\rm eff}(x)\psi_{\rm
  eff}(x)=E\psi_{\rm eff}(x),\label{seq} 
\end{equation}
or 
\begin{equation}
 V_{\rm eff}(x)=E+{1\over 2}{\psi_{\rm eff}''(x)\over\psi_{\rm
 eff}(x)}.
\end{equation}
In terms of $\rho(x)$ this translates into
\begin{equation}
 V_{\rm eff}(x)=E+\frac{1}{4}\left[{\rho''(x)\over\rho(x)}-{1\over
     2}\left({\rho'(x)\over\rho(x)}\right)^2\right].
\label{pot} 
\end{equation}
By construction this potential is such that the ground state wavefunction can be used to
reproduce the NRG derived $\rho(x)$.

\subsection{Results for the symmetric model}

Unless otherwise stated we use in the following the parameters $W=2D=2$,
$\pi\Delta=0.1$ and $\omega_0=0.1$ setting the energy scales for electrons and
phonons in this section. The 
phonon frequency $\omega_0$ has been chosen so that $1/\omega_0$ is on the
scale of the life time of a electron on the impurity site, $\omega_0\sim
\pi\Delta$. We know in the adiabatic limit $\omega_0\ll \pi\Delta$, the $x$
coordinate becomes a classical variable, and in the opposite limit
$\omega_0\to\infty$ it maps on to an effective Hubbard model, so the range
$\omega_0\sim\pi\Delta$ is the interesting 
one to investigate.  

\begin{figure}[!t]
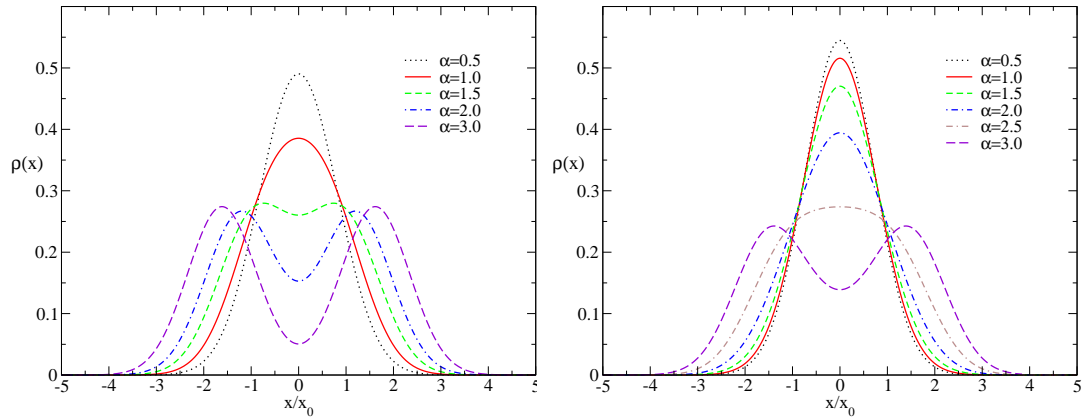

\centering
 \includegraphics[width=0.45\textwidth]{figures_alex/rhox1.eps}
\vspace*{0.8cm}
 \includegraphics[width=0.45\textwidth]{figures_alex/rhox2.eps}
      \caption{The total probability distribution function $\rho(x)$ for the
    oscillator displacement $x$ in the ground state for a range of values of
   $\alpha$, in the left panel with $U=0$ and the right
with $U/\pi\Delta=2$.}
    \label{rhox1}
\end{figure}

In  Fig. \ref{rhox1} we give results for the probability distribution 
function $\rho(x)$ for the symmetric model calculated as outlined above.

\noindent
The results on the left hand panel correspond to $U=0$
($\epsilon_d=0$). We take as a relative measure of the strength of the
phonon coupling and hybridisation scale the dimensionless factor
$\alpha=\lambda/\pi\Delta$ ranging from weak ($\alpha\ll 1$) to strong
coupling ($\alpha\gg 1$).  As $\alpha$ increases the distribution broadens,
and for  $\alpha=1.5$ a  two peak  structure can be seen which becomes more 
marked  as the coupling strength is increased further. 

\begin{figure}[!htbp]
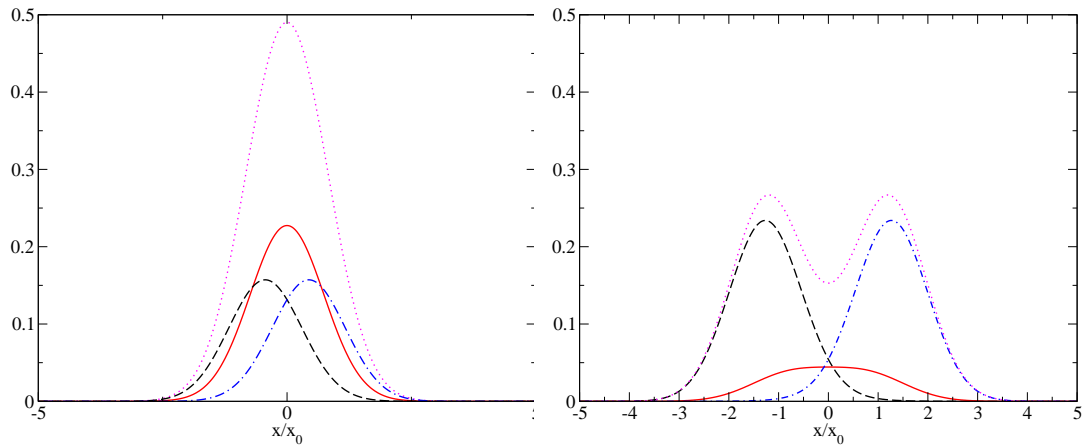

  \begin{center}
    \includegraphics[width=0.45\textwidth]{figures_alex/rho_0.5.eps}
 \includegraphics[width=0.45\textwidth]{figures_alex/rho_2.eps}
    \caption{The components, $\rho(x:0,0)$ (dot-dashed curve),
      $\rho(x:1,1/2)+\rho(x:1,-1/2)$ (full curve)
and  $\rho(x:2,0)$ (dashed) of $\rho(x)$ (dots) for the case $\alpha=0.5$
(left panel), and $\alpha=2.0$ (right panel) for $U=0$.} 
    \label{rho_2}
  \end{center}
\end{figure}
In Fig. \ref{rho_2} we plot the individual components of $\rho(x)$;
$\rho(x:0,0)$,  $\sum_{\pm}\rho(x:1,\pm 1/2)$ and  $\rho(x:2,0)$,
corresponding to $q=0,1,2$,
for  $U=0$, for the two cases $\alpha=0.5$ (left panel)
and $\alpha=2$ (right panel). One can see from these curves the two factors
that lead to the two peak structure on increasing $\alpha$. One factor is
that the maxima of curves corresponding to $q=0$ and $q=2$ are shifted on
either side of central peak corresponding to $q=1$. This reflects that fact
that in these charge
states for an
isolated
impurity  the oscillator is displaced from $x=0$
to $\sqrt{\lambda}/\omega_0$ and  $-\sqrt{\lambda}/\omega_0$ respectively.
 The other factor is
that the  weights of the peaks at  $q=0$ and $q=2$,  compared to the weight of the central
peak corresponding to $q=1$, increases with $\alpha$. The integrated weight under the
curve $P_q$ is a measure of the probability of the occupation of the local
level in the state with charge quantum number $q$.
This shift in relative weights is due to the fact that the coupling to the
phonon mode induces a local attraction. The weight $P_2$ measures the
probability of the local level being doubly occupied which is equal to the expectation
value $\langle n_{f,\uparrow}n_{f,\downarrow}\rangle$. For $\alpha=2$, the weights
$P_0=P_2=0.430261$ are in precise six figure agreement with $\langle
n_{f,\uparrow}n_{f,\downarrow}\rangle$ as calculated directly from the NRG
calculation, and we have $P_1=1-P_0-P_2=0.139478$. Notice that for $\alpha=0$  the corresponding values 
are $P_0=P_2=0.25$,
$P_1=0.5$. The relative shift of the weights can be explained to a large extent, but
not completely, by the local induced attractive interaction.
If we take a model with a local  attractive interaction
$U/\pi\Delta=-2$, but no phonon coupling, then the values deduced from the NRG
for $P_q$ are  $P_0=P_2=0.38289$, $P_1=0.23422$, which underestimates 
the relative shift away from the state $q=1$ found in the phonon
coupled model with $\alpha=2$. This shows that the energetic gain of creating
zero and doubly occupied sites is higher in the system with phonons as
compared to the case with instantaneous attraction.

 \begin{figure}[!htbp]
   \begin{center}
     \includegraphics[width=0.45\textwidth]{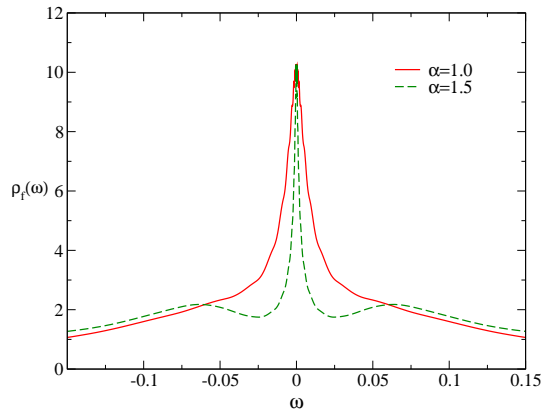}
     \caption{The spectral density $\rho_f(\omega)$ for the impurity single
       electron Green's function for $U=0$ and $\alpha=1,1.5$.}
     \label{rhof}
   \end{center}
 \end{figure}
\noindent

The  spectral density for the local f-electron  Green's function
$\rho_f(\omega)$ is shown in Fig. \ref{rhof} for $U=0$ for $\alpha=1$ and
$\alpha=1.5$. It can be seen that the development of the two peak form in
$\rho(x)$ correlates with the development of three peaks in
$\rho_f(\omega)$. Clear shoulders can be seen in the result for
$\rho_f(\omega)$  for $\alpha=1$, and the three peak form has fully emerged
for $\alpha=1.5$. The central peak becomes extremely narrow for values of
$\alpha$ in the strong coupling regime \cite{HM02}. 

\begin{figure}[!htbp]
  \begin{center}
    \includegraphics[width=0.45\textwidth]{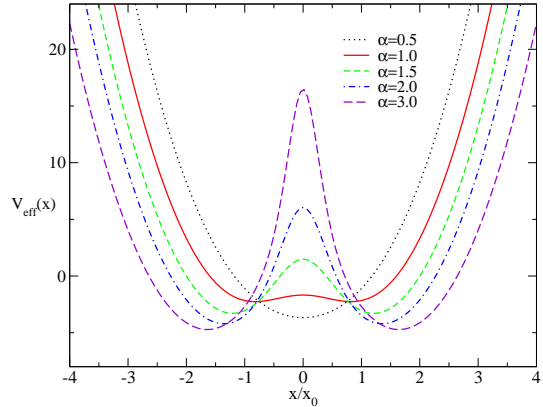}
    \caption{The  effective potential $V_{\rm eff}(x)$ for $U=0$ and  a range of values of $\alpha$,
      corresponding to the results for $\rho(x)$ in the left panel of figure
      \ref{rhox1}.} 
    \label{pot1}
  \end{center}
\end{figure}
The effect of the interaction $U$ in suppressing 
the onset of the double peak distribution can be seen by comparing the $U=0$
results with those shown in the right panel of  Fig. \ref{rhox1}
which are for the case $U/\pi\Delta=2$ ($\epsilon_f=-U/2$).   
The double peak structure, which develops in the $U=0$ case for $\alpha\sim
1.5$, develops in the  $U/\pi\Delta=2$ case  only when $\alpha\sim 2.5$. This
is a smaller value than what is expected naively in the $\omega_0\to \infty$ case
where $U_{\rm eff}=U-\lambda$ signaling again that for finite $\omega_0$ when phonons can be excited
the occupation of the  zero and doubly occupied sites is energetically
more favourable. 
\par 

The corresponding effective potentials $V_{\rm eff}(x)$ for the $U=0$ case, as
deduced from  equation (\ref{pot}), are shown in Fig.
\ref{pot1}. The on-set of a double well feature can be seen in the results
for  $\alpha=1.0$, which is before a double peak structure can be seen in the
corresponding $\rho(x)$. The effective potentials for higher values of
$\alpha$ have a clear double well form. 

It is of interest to compare this potential with one calculated using a semiclassical
approximation, in which we neglect the kinetic energy of
the oscillator and treat the coordinate $x$ as a classical variable. This is
a commonly used approximation in taking the electron-phonon coupling into
account, and corresponds to the Born-Oppenheimer approximation.
Evaluating the impurity contribution to the total ground state energy $E(x)$ as a function of $x$ one
arrives at an expression for the effective semiclassical potential, $V_{\rm
  s-cl}(x)$ ($=E(x)$) given by 
\begin{equation}
V_{\rm s-cl}(x)=\epsilon_f-\frac{2\Delta}{\pi}+\frac{\omega_0^2x^2}{2}
-\frac{2\bar\epsilon_f(x)}{\pi}{\rm tan}^{-1}\Big(\frac{\bar \epsilon_f(x)}{\Delta}\Big)
+\frac{\Delta}{\pi} \log\Big(\frac{\bar \epsilon_f^2(x)+\Delta^2}{D^2}\Big),
\label{impVscl}
\end{equation}
where $\bar\epsilon_f(x)=\epsilon_f+\sqrt{2\omega_0}gx$.

\begin{figure}[!htbp]
  \begin{center}
    \includegraphics[width=0.45\textwidth]{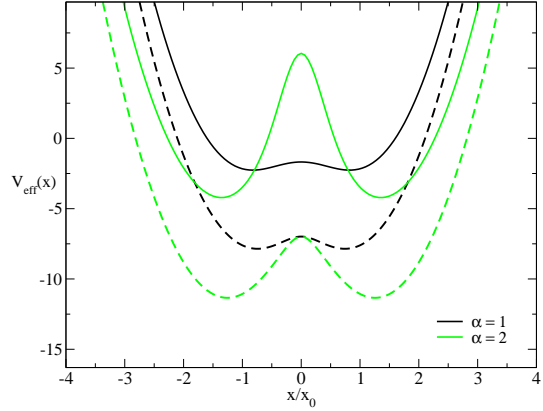}
    \caption{The effective potentials $V_{\rm eff}(x)$ (full lines) as
      deduced from equation (\ref{pot}), compared with corresponding the
      semiclassical potentials $V_{\rm s-cl}(x)$ (dashed lines), Eq. (\ref{impVscl}),
      for $U=0$.} 
    \label{pot2}
  \end{center}
\end{figure}

 In Fig. \ref{pot2} we compare the semiclassical potential $V_{\rm s-cl}(x)$  with $V_{\rm
  eff}(x)$ deduced from equation  (\ref{pot})  for
$\alpha=1$ and $\alpha=2$. 

\noindent

The relevant comparison is in the shapes of these
potentials, not their absolute values, and they have been subject to a
constant shift so their forms can be compared more easily. It can be seen that
the potentials develop in a similar way as the coupling strength is increased,
though when the double well form develops the potential barrier between the
wells is less marked in   the semiclassical case.  The minima  occur at very
similar $x/x_0$-values; 
 $-0.74$ ($V_{\rm s-cl}(x)$)  and $-0.79$ ($V_{\rm eff}(x)$)  for $\alpha=1$, and 
$-1.26$ ($V_{\rm s-cl}(x)$)  and $-1.35$ ($V_{\rm eff}(x)$)  for
$\alpha=2$.\par The coefficient of the $x^2$ term in $V_{\rm s-cl}(x)$ changes
sign for $\alpha=0.5$, which is the point at which the double well begins to
form.  For $\alpha>0.5$ there are two mean field solutions for the expectation
value $\langle \hat x \rangle$ corresponding to the two minima in  $V_{\rm
  s-cl}(x)$ and the local symmetry is broken. The positions of the minima in
$V_{\rm s-cl}(x)$  can be deduced from the equation,    
\begin{equation}
\frac{\partial V_{\rm s-cl}(x)}{\partial x}=0=\omega_0^2 x
+\sqrt{2\omega_0}g(n_f-1),
\end{equation}
where $n_f$ is the mean field occupation value given by
\begin{equation}
n_f=1-{2\over\pi}{\rm
    tan}^{-1}\left(\frac{\bar \epsilon_f(x)}{\Delta}\right),
\end{equation}
yielding 
\begin{equation}
x =\langle \hat x\rangle_{\rm  MF}=-{\sqrt{\lambda}\over \omega_0}( n_f-1).
\label{sclx}\end{equation}
We can deduce an exact relation of this type  by  introducing an additional to term of the
form $c(b+b^\dagger)$  into the Hamiltonian. The ground state energy $E(c)$
will be a function of the coupling $c$ introduced, and we can deduce that
\begin{equation}
\left.{\partial E(c)\over\partial c}\right|_{c=0}=\langle b+b^\dagger\rangle=
\sqrt{2\omega_0}\langle \hat x \rangle.
\label{ec}
\end{equation}
If we now  perform a canonical transformation $H'=\hat U^{-1}H\hat U$ with 
$\hat U=e^{-c(b^\dagger -b)/\omega_0}$, the terms in $H'$ which depend on $c$
are 
\begin{equation}
-{c^2\over\omega_0}-{2gc\over\omega_0}(\langle \hat n_f\rangle -1).
\end{equation} 
As the canonical transformation does not effect the energy values we can
use this result in equation (\ref{ec}) to determine $\langle
b+b^\dagger\rangle$, which leads to the result,
 \begin{equation}
\langle \hat x\rangle=-{\sqrt{\lambda}\over \omega_0}(\langle \hat
n_f\rangle-1).
\label{x}
\end{equation} 
This is the same formula as derived from the semi-classical approximation as
given  in
equation (\ref{sclx}), 
except that the mean field value $n_f$ is replaced by the exact value  $\langle \hat n_f\rangle$
for the impurity occupation. This result  holds for $U\ne 0$ and is exact.

In the mean field broken symmetry solutions we have the two broken symmetry
solutions at strong coupling, such that   $n_f\sim 0$ and $n_f\sim 2$, giving
from equation  (\ref{sclx}) $\langle \hat x\rangle_{\rm  MF}\sim
\pm{\sqrt{\lambda}/\omega_0}$ for the positions of the two minima $V_{\rm
  s-cl}(x)$. In the exact solution for the impurity model, however, this
broken symmetry must be  restored and the average value of $x$ must be
zero. This is clearly the case in the NRG solution as $\langle \hat x \rangle$
can be deduced from $\rho(x)$, using $\langle \hat x
\rangle=\int_{-\infty}^{\infty} x\rho(x)\, dx$, and is zero as  $\rho(x)$ is
symmetric. However we have seen 
that the positions of the minima in both $V_{\rm eff}(x)$ and $V_{\rm
  s-cl}(x)$ are very close so the mean field values for $\langle \hat
x\rangle_{\rm MF}$ provide an estimate of the positions of the minima in
$V_{\rm eff}(x)$. 

\begin{figure}[!htbp]
  \begin{center}
    \includegraphics[width=0.45\textwidth]{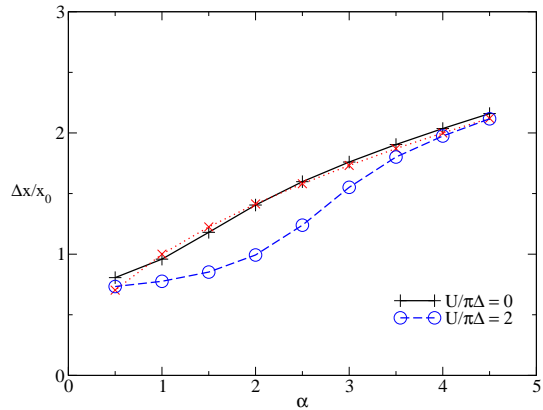}
    \vspace*{0.2cm}
    \caption{The root mean deviation $\Delta x/x_0$  as a function of $\alpha$ for the case $U=0$,
      $\epsilon_f=0$ (full curve), and  $U/\pi\Delta=2$, $\epsilon_f=-U/2$
      (dashed curve). The dotted curve corresponds to $\sqrt{\alpha\,\pi\Delta}/\omega_0$. }
    \label{sym}
  \end{center}
\end{figure}

As $\langle \hat x\rangle=0$ in the exact solution for the symmetric model,
a more interesting quantity to calculate is the root mean square deviation
$\Delta x$, where $ (\Delta x)^2=\langle (\hat x-\langle \hat x\rangle)^2\rangle$. In
Fig. \ref{sym} we give a plot of $\Delta x$ deduced from $\rho(x)$ for the
model with $U=0$ and several values of $\alpha$.  

\noindent 

It can be seen that $\Delta x$ does vary significantly with $\alpha$
reflecting the fact that in the strong coupling regime the deviation $\Delta
x$ is determined  by position of the minima in the double potential wells. As
the mean field values $\langle \hat x\rangle_{\rm MF}$, were found to give a good
estimate of these, and in the strong coupling regime  $\langle \hat x\rangle_{\rm 
  MF}\sim \pm{\sqrt{\alpha\,\pi\Delta}/\omega_0}$, we compare in Fig. \ref{sym}  this estimate with the
calculated value of $\Delta x$. It is seen to provide a good estimate of 
 $\Delta x$ over the range of $\alpha$ shown.
 Also in Fig \ref{sym} we give the values of $\Delta x$ for the symmetric
 model with $U/\pi\Delta=2$. The tendency for the $U$ term to suppress the
 fluctuations for the lower values of the coupling strength $\lambda$, noted
 in Fig. \ref{rhox1}, is apparent but to have only a relatively minor effect
 in the stronger coupling range.\par

\subsection{Results for the asymmetric model}
\begin{figure}[!htbp]
  \begin{center}
    \includegraphics[width=0.45\textwidth]{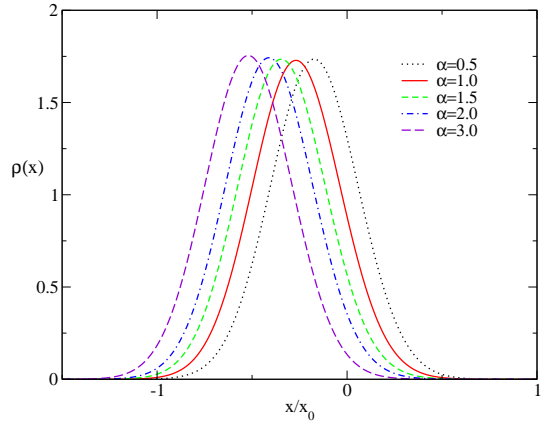}
    \vspace*{0.2cm}
    \caption{The total probability distribution function $\rho(x)$ for the
      oscillator displacement $x$ in the ground state for a range of values of
      $\alpha$ with $U=0$ and $\epsilon_f/\pi\Delta=-0.5$. }
    \label{asymrho}
  \end{center}
\end{figure}

With only a small degree of asymmetry, the form for $\rho(x)$ changes
quite dramatically in the strong interaction regime. This is because the
doubly occupied impurity state is  now predominantly favoured.
In Fig. \ref{asymrho} we show results for $\rho(x)$ in a case
with $\epsilon_f/\pi\Delta=-0.5$ and $U=0$ for the same range of values of
$\alpha$ as in Fig. \ref{rhox1}.  

\begin{figure}[!htbp]
  \begin{center}
    \includegraphics[width=0.45\textwidth]{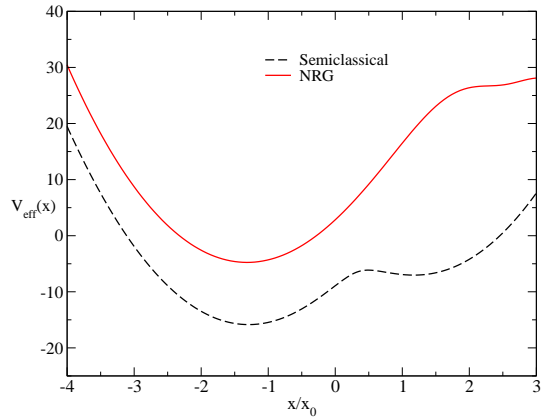}
    \vspace*{0.2cm}
    \caption{The effective potentials $V_{\rm eff}(x)$ (full curve) as deduced
      from equation (\ref{pot}), compared with corresponding the semiclassical
      potentials $V_{\rm s-cl}(x)$ (dashed curve) for  $\alpha=2.0$,
      $\epsilon_f/\pi\Delta=-0.5$ and $U=0$.}
    \label{pot3}
  \end{center}
\end{figure}

\noindent
There is just a single narrow peak for $\rho(x)$ in each case
 which shifts to slightly more negative values of $x$ as $\alpha$ increases.
In Fig. \ref{pot3} we compare the semiclassical potential and $V_{\rm
  eff}(x)$ as derived from equation (\ref{pot}) for the case $\alpha=1$. 

\begin{figure}[!htbp]
  \begin{center}
\includegraphics[width=0.45\textwidth]{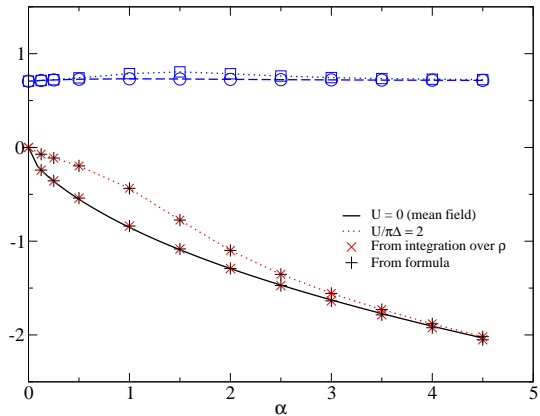}
    \vspace*{0.2cm}
    \caption{The values of $\langle \hat x\rangle/x_0$ for $U=0$, $\epsilon_f/\pi\Delta=-0.5$,
      and  $U/\pi\Delta=2$, $\epsilon_f=-U/2-0.05$, as 
      calculated from the average of $\rho(x)$ (crosses)
      and those deduced from equation (\ref{x}) (plus signs). The mean field
      results for $U=0$ correspond to the full curve. The corresponding
      results for the  root mean deviation
      $\Delta x/x_0$  are also shown, for the case $U=0$, $\epsilon_f/\pi\Delta=-0.5$,
      (dashed curve and circles), and  $U/\pi\Delta=2$, $\epsilon_f=-U/2$
      (dotted curve and squares).}
    \label{asym1}
  \end{center}
\end{figure}

\noindent
Both potentials have an 
absolute minimum at $x/x_0=-1.29$ though in the semiclassical case there is
a secondary local minimum. This value of $x/x_0$ agrees with that predicted
by equation (\ref{x}) using the value derived from $\langle\hat n_f\rangle$
derived from the NRG calculation which gives $x/x_0=-1.2909$.
 We can check the relation (\ref{x}) by
taking the average of $x$ over the distribution $\rho(x)$ and then compare with the
result  from  equation (\ref{x}) using the NRG calculated $\langle\hat
n_f\rangle$. The results for a range of values of $\alpha$ are shown
 as points (crosses and plus signs) in Fig. \ref{asym1} (smaller values for $U=0$, $\epsilon_f/\pi\Delta=-0.5$,
      and  larger values $U/\pi\Delta=2$, $\epsilon_f=-U/2-0.05$). 

\noindent
For both sets of parameters, the results of the two calculations are in
remarkably good agreement, giving the  same values to at least five
significant figures in all cases. The full curve in Fig. \ref{asym1}
corresponds to the mean field result ($U=0$) for $\langle 
\hat x\rangle/x_0$ and can be seen also to be in good agreement with the exact results.
 Also shown in Fig. $\ref{asym1}$ is the root mean square $\Delta x$ calculated from
$\rho(x)$.  Though the average displacement $\langle \hat x\rangle$ increases with
increasing  $\alpha$ it can be seen that  $\Delta x$ remains almost constant.  
In mean field theory $\Delta x=0$, as in this approximation $\langle
\hat x^2\rangle=\langle \hat x\rangle^2$. In the semiclassical approach one could
estimate $\rho(x)$ by solving the Schr\" odinger equation (\ref{seq})
with the potential $V_{\rm s-cl}(x)$ and using  $\rho(x)=|\psi_{\rm
  gs}(x)|^2$, where $\psi_{\rm  gs}(x)$ in the ground state wavefunction,
and then use the result to take an average of $x^2$. We can, however, 
calculate it exactly in the limit of very weak and strong coupling limits.
In the uncoupled case using the ground state wave function for the oscillator
we find $(\Delta x)^2/x_0^2=1/2$. In the strong coupling case with
asymmetry we can take $\langle \hat n_f\rangle =2$ or $\langle \hat
n_f\rangle=0$, and use a displaced oscillator transformation to new phonon
creation and annihilation operators,  
 $a^{\dagger}$, $a^{(\dagger)}=b^{(\dagger)}\pm g/\omega_0$. The ground state
 $|{\rm gs}\rangle$ then
corresponds to the state $a|{\rm gs}\rangle=0$, and in this state $\langle
b^{\dagger}b\rangle=g^2/\omega_0^2$,  $\langle
 \hat x^2\rangle=x_0^2(1+4g^2/\omega_0^2)/2$. We have from equation
(\ref{x}) with   $\langle (\hat n_f-1)\rangle =\pm 1$, $\langle
x\rangle^2=2g^2x_0^2/\omega_0^2$, which gives again $(\Delta x)^2/x_0^2=1/2$,
so that we find  $\Delta x/x_0=1/\sqrt{2}$ in both limits. This agrees well
with the results shown in Fig. \ref{asym1} and  $\Delta x$ does not deviate
much from this value over the whole range of $\alpha$. As will be seen in the
Sec. \ref{HHmodel} for the case of the lattice model,  the results will be
different for $\Delta x$ near the transition to a charge ordered state.
\par
The corresponding values for the case with $U/\pi\Delta=2$,
$\epsilon_f/\pi\Delta=-0.15$ are shown in the same Fig. \ref{asym1}. The effect of 
finite $U$ is to suppress the value of $\langle \hat x\rangle$ for smaller values
of $\alpha$, but only has a marginal effect for $\alpha\ge 3$ and has very
little effect on $\Delta x$. Again there is  five figure agreement in the 
the two calculations for $\langle  \hat x\rangle$; the one
based on integrating over $\rho(x)$ and the one using Eq. (\ref{x}).

\begin{figure}[!htbp]
  \begin{center}
    \includegraphics[width=0.45\textwidth]{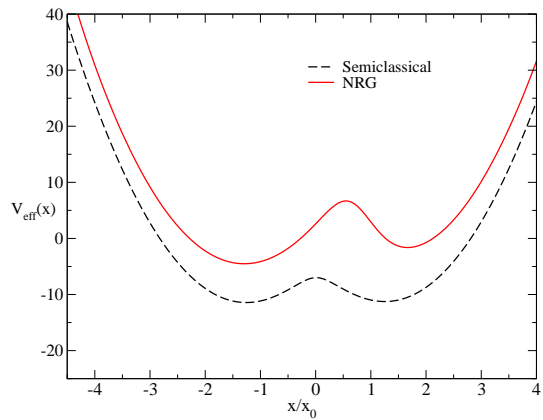}
    \caption{The effective potentials $V_{\rm eff}(x)$ (full curve) as deduced
      from equation (\ref{pot}), compared with corresponding the semiclassical
      potentials $V_{\rm s-cl}(x)$ (dashed curve) for  $\alpha=2.0$,
      $\epsilon_f/\pi\Delta=-0.01$ and $U=0$. }
    \label{pot4}
  \end{center}
\end{figure}

As the model in the limit $\omega_0\to\infty$ ($\lambda$ finite)
corresponds to an Anderson model with an interaction
term $U-\lambda$. For $U=0$ and finite $\lambda$, therefore, it becomes
equivalent to a negative-$U$ Anderson model. The symmetric model  in the regime
$\lambda/\pi\Delta \gg 1$  has a Kondo effect due to charge
rather than spin fluctuations. Introducing some asymmetry by changing
$\epsilon_f$ from the value for the symmetric case is equivalent to
introducing a magnetic field in the Kondo case \cite{HBK06}, which for large
fields suppresses the Kondo effect. We found that in using a value
$\epsilon_f/\pi\Delta=-0.5$  that the 
mean field estimate for $\langle \hat x\rangle$ and the exact result were in
good agreement. This is due to the fact that this degree of asymmetry
corresponds to the Kondo case with a large magnetic field, and also because the value of
 $\omega_0$ used is much smaller than the bandwidth $D$.
For a much smaller degree of asymmetry, which would correspond to a smaller
'magnetic' field, we should expect to find some
limitations in the predictions from the semiclassical approximation.
To examine this further we consider the case with
$\alpha=2$ and $\epsilon_f/\pi\Delta=-0.01$. In Fig.
\ref{pot4} we show the semiclassical potential and $V_{\rm eff}(x)$ derived
from (\ref{pot}).  

\noindent
In this case we see that both potentials have two local minima. The absolute
minimum in the two cases coincide at a value of $x/x_0=-1.26$, which
corresponds to the mean field estimate of $\langle \hat x\rangle/x_0$. The
value obtained by averaging $x$ over 
the distribution $\rho(x)$, and from the formula (\ref{x}) with the NRG value
for $\langle \hat n_f\rangle$, both give $\langle \hat x\rangle/x_0=-1.09888$.
The average value in this case  no longer coincides with the absolute
minimum of the potential as the fluctuations to the local
neighbouring  minimum make a significant contribution. As the semiclassical
 equation (\ref{sclx})  for $\langle \hat x\rangle$ agrees with the exact one
in equation (\ref{x}), this difference arises from the fact that the
semiclassical
prediction for $n_f$ disagrees with the exact value of  $\langle \hat
n_f\rangle$. The semiclassical prediction gives $n_f=1.889$ and the
exact value from the NRG gives $\langle \hat n_f\rangle=1.777$,
which explains the difference in the predictions. It is interesting to note,
however, that the semiclassical prediction does coincide with the absolute
minimum of the effective potential derived from the NRG results.\par
\begin{figure}[!htbp]
  \begin{center}
\includegraphics[width=0.45\textwidth]{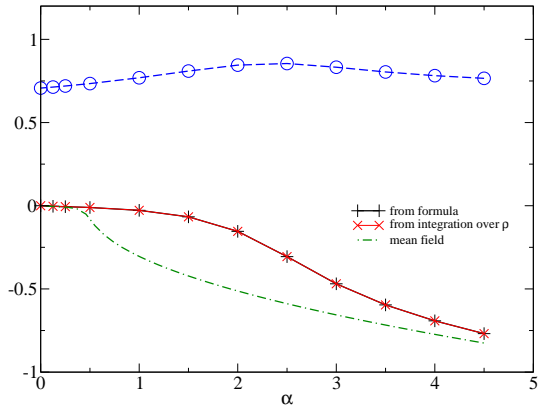}
    \vspace*{0.2cm}
    \caption{The average oscillator displacement $\langle \hat x\rangle/x_0$
      (full curve) as a function of $\alpha$ for the case $U=0$,
      $\epsilon_f/\pi\Delta=-0.01$, $\omega_0=0.6$. The values calculated from the average
      over $\rho(x)$ are indicated by crosses and those deduced  from equation
      (\ref{x}) by plus signs. The dashed-dot curve gives the corresponding
      mean field results. Also shown are the results of the root mean square
      deviation $\Delta x/x_0$ (dashed curve and circles).}
    \label{asym2}
  \end{center}
\end{figure}

The deviations from mean field theory become more marked for higher oscillator
frequencies. In Fig. \ref{asym2} we give a plot of $\langle x\rangle/x_0$
for the case $U=0$ and $\epsilon_f/\pi\Delta=-0.1$ taking the phonon frequency
value $\omega_0=0.6$. 

\noindent
In this case, except for the small
values of $\alpha$, there is quite a discrepancy between the exact and mean
field results. The corresponding results for the root mean square deviation
$\Delta x$ are also shown. The results for this quantity are very similar to
those shown in Fig. \ref{asym1}. 

\begin{figure}[!htbp]
  \begin{center}
    \includegraphics[width=0.45\textwidth]{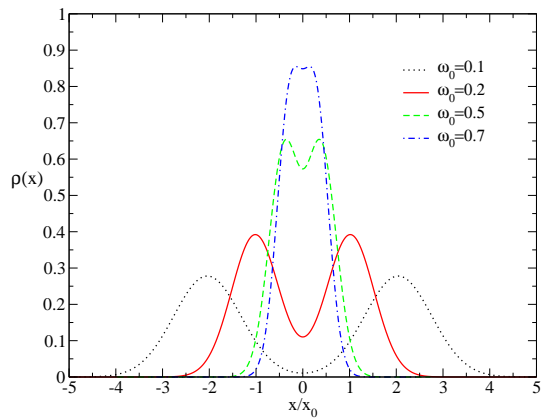}
    \caption{The variation of $\rho(x)$ with the frequency $\omega_0$ in the
      strong coupling case with $\alpha=4.5$. The value of $x_0$ is fixed and
      corresponds to $1/\sqrt{\omega_0}$ for $\omega_0=0.1$.} 
    \label{wrhox}
  \end{center}
\end{figure}

We can also examine the dependence of $\rho(x)$ on the oscillator frequency
$\omega_0$.  In Fig. \ref{wrhox} we show the change in form of $\rho(x)$ as
the frequency is increased   for a fixed value of $\lambda$ in the strong
coupling regime corresponding to $\alpha=4.5$ ($U=0$, $\epsilon_f=0$). 

\noindent
We argued earlier that the minimum in the effective potential in the strong
coupling regime occurs at a value of $x\sim \sqrt{\lambda}/\omega_0$. We would
expect the peak in $\rho(x)$ to behave in a similar way, so for fixed
$\lambda$  the peak positions should vary as $1/\omega_0$. This can be seen to
be well satisfied in the results shown in Fig. \ref{wrhox}. In the limit
$\omega_0\to \infty$ the double peak feature disappears entirely and $\rho(x)$
becomes a delta function at $x=0$. In this limit the mean field equation for
$n_f$ still has a broken symmetry solution for $\alpha>0.5$. This  coincides
with the static mean field solution for the Anderson model with  $U=-\lambda$,
if one first performs a Hubbard-Stratonovich transformation to couple the auxiliary
field $x(\tau)$ solely to the impurity charge.
It differs by a factor of 2 from the mean field theory of the Anderson
model with  $U=-\lambda$, where the interaction term
$Un_{d,\uparrow}n_{d,\downarrow}$ is approximated by $U(n_{d,\uparrow}\langle
n_{d,\downarrow}\rangle+n_{d,\downarrow}\langle n_{d,\uparrow}\rangle-
\langle n_{d,\uparrow}\rangle \langle n_{d,\downarrow}\rangle) $.

 \section{Local Fluctuations in the  Holstein-Hubbard  Model}
\label{HHmodel}
The states of broken symmetry predicted by
the semiclassical/mean field theory
for larger values of $\lambda$ cannot exist for the impurity
Anderson-Holstein model; the symmetry has to be restored in the exact
solution. Symmetry breaking, however, as a result of a phase transition  can
occur in a lattice model. To study $\rho(x)$ in the neighbourhood of a phase
transition we consider the Holstein-Hubbard model, described by the
Hamiltonian, 
\begin{eqnarray}
  \label{hubholham}
  H&=&\sum_{i,j,{\sigma}}(t_{ij}\elcre i{\sigma}\elann
j{\sigma}+\hc)+U\sum_i\hat n_{i,\uparrow}\hat n_{i,\downarrow} \\
&&+\omega_0\sum_ib_i^{\dagger}b_i+g\sum_i(b_i+b_i^{\dagger})\Big(\sum_{\sigma}\hat
n_{i,\sigma}-1\Big).
\nonumber
\end{eqnarray}
$\elcre i{\sigma}$ creates an electron at lattice site $i$ with spin $\sigma$,
and $b_i^{\dagger}$ a phonon with oscillator frequency $\omega_0$,
$\hat n_{i,\sigma}=\elcre i{\sigma}\elann 
i{\sigma}$. There is a coupling $g$ to the local charge at each site, as in
the Holstein model, and an  on-site interaction $U$ between spin-up and
spin-down electrons, as in the Hubbard model. The  hopping term $t_{ij}$
between orbitals localised on each site leads to a conduction band with a
density of states $D_0(\omega)$ when $g=U=0$. In the limit of infinite
dimensions the model can be mapped into an effective Anderson-Holstein
model, which can then be solved using the NRG. This has been described
fully elsewhere \cite{GKKR96,BCP08}, and is known as the dynamical mean field
theory (DMFT). For  three dimensional systems, the mapping is only
approximate. This approach is non-perturbative and can be applied in the strong interaction regime
of these models to describe strong correlation effects, and indications are that it
constitutes a good approximation when the self-energy of the electrons is local and a function of
frequency only.\par
There have been several applications of the DMFT method to study phase
transitions in the Hubbard-Holstein model \cite{BZ98,MHB02,KMH04,KHE05}.  There are various possible
transitions to states of broken symmetry in this model;  bipolaronic  (BP),
charge ordered  (CO), antiferromagnetic (AFM) and the superconducting (SC)
state.
We restrict our attention here to the case of half-filling, so we do not
include the superconducting case, which exists as a stable state
away from half-filling. The transition first studied by the DMFT-NRG
method for this model did not include the possibility of either charge order
or antiferromagnetism \cite{MHB02,KMH04}.  There is, however, a metal-insulator transition from the
normal state (N) to the bipolaronic state (BP), first studied for the model
with $U=0$, which occurs as the
electron-phonon coupling $\lambda$ is increased at a critical value
$\lambda_c$. The transition also occurs in the model with $U\ne 0$,
at larger values of $\lambda_c$, as the attractive term induced by
$\lambda$ has to overcome the repulsion due to $U(>0)$. If the possibility of
transitions to charge order (CO)
and antiferromagnetism are included, then it has been found that
antiferromagnetism occurs for $U-\lambda>0$ and charge order for
$U-\lambda<0$ \cite{Bau09pre,BH09pre}.

For the  DMFT-NRG calculations presented here we have taken a Bethe lattice
form for the density of states $D_0(\omega)$ of the conduction electrons, 
\begin{equation}
D_0(\omega)=  {1\over 2\pi t^2}\sqrt{4t^2-\omega^2}.
\end{equation}
We choose a value $t=1$ to set the energy scale in the following, which
corresponds to a bandwidth $W=4t=4$. 
The physically relevant regime in the lattice case is for phonon frequencies
small compared with the bandwidth. In most of the calculations in this section
we take $\omega_0=0.6$, which is small compared with the bandwidth of $W$ but
well away form the adiabatic limit.  The distribution function for the
local oscillator displacement $\rho(x)$ was calculated from the DMFT-NRG
density matrix using equation (\ref{rho}).

\subsection{Model without long range order}
\begin{figure}[!htbp]
\centering
\includegraphics[width=0.45\textwidth]{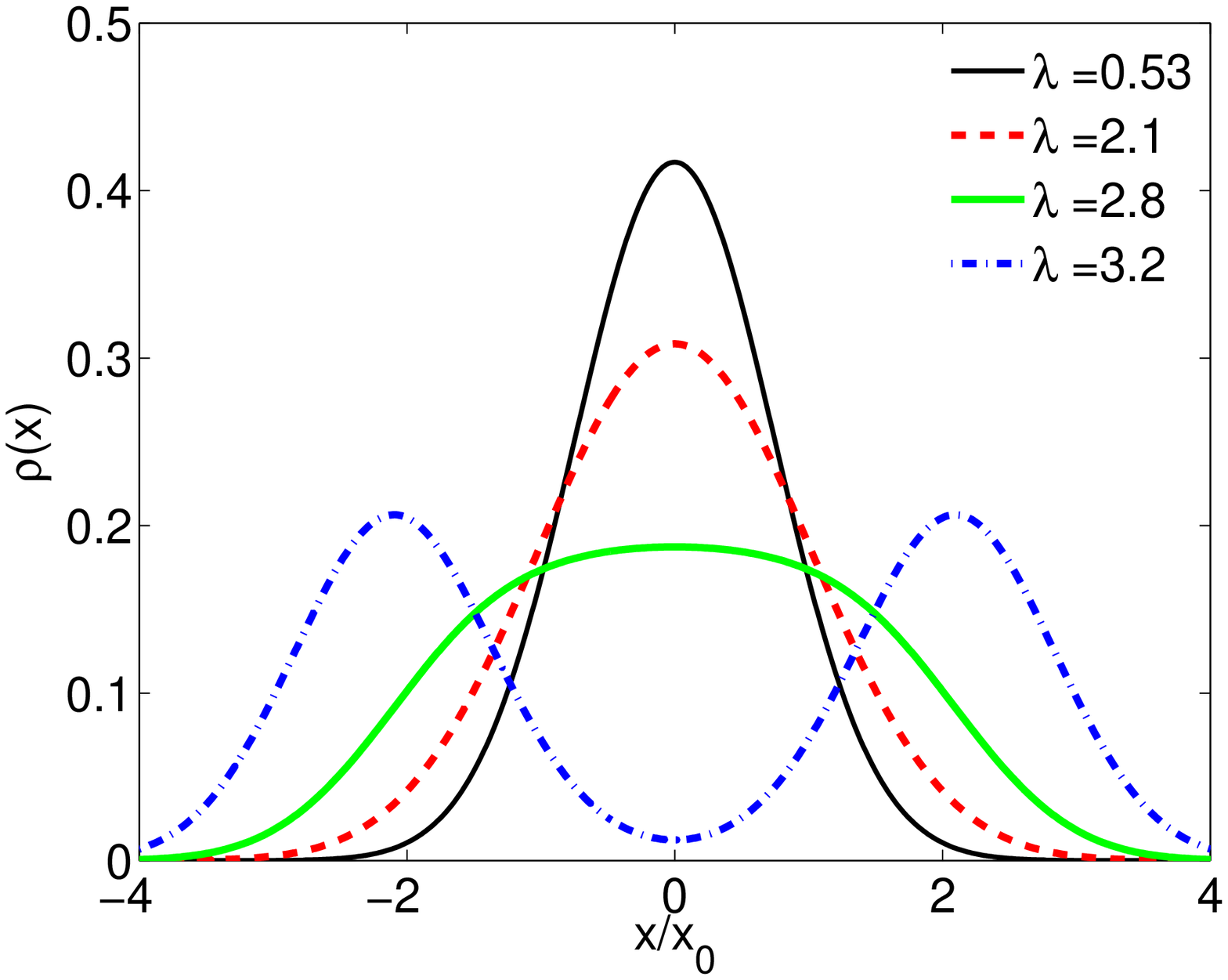}
\includegraphics[width=0.45\textwidth]{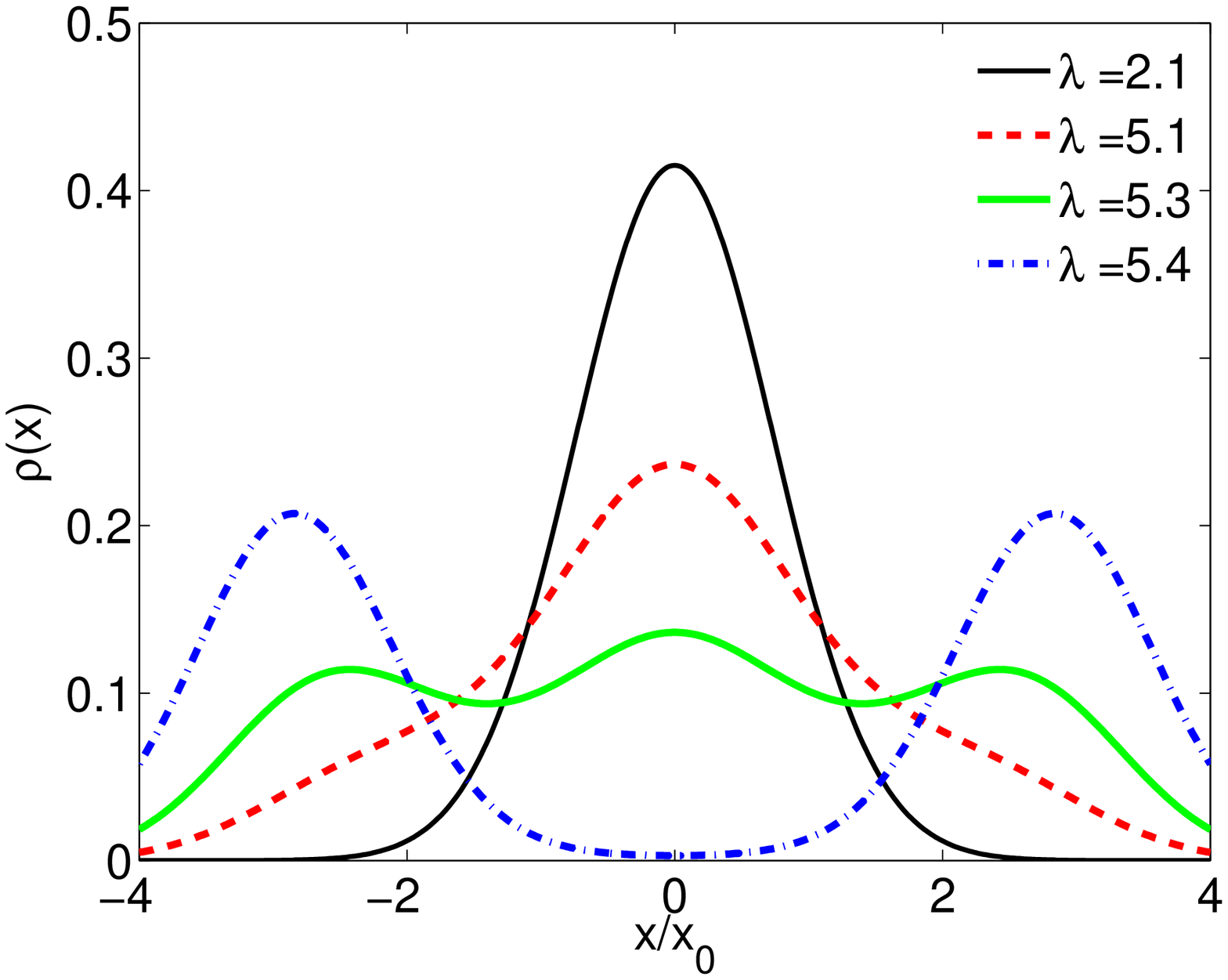}
\caption{(Color online) The local probability distribution $\rho(x)$ for $U=2$
  (right part) for the N state and $U=5$
  (left part) for various values of $\lambda$ for $\omega_0=0.6$.}       
\label{rhox_xdepvarlambdaN}
\end{figure}

We consider first the results for the normal to bipolaronic transition, which
are shown in Fig. \ref{rhox_xdepvarlambdaN} for $U=2$ (left panel) and  
$U=5$ (right panel) and different values of $\lambda$.

\begin{figure}[!htbp]
\centering
\includegraphics[width=0.45\textwidth]{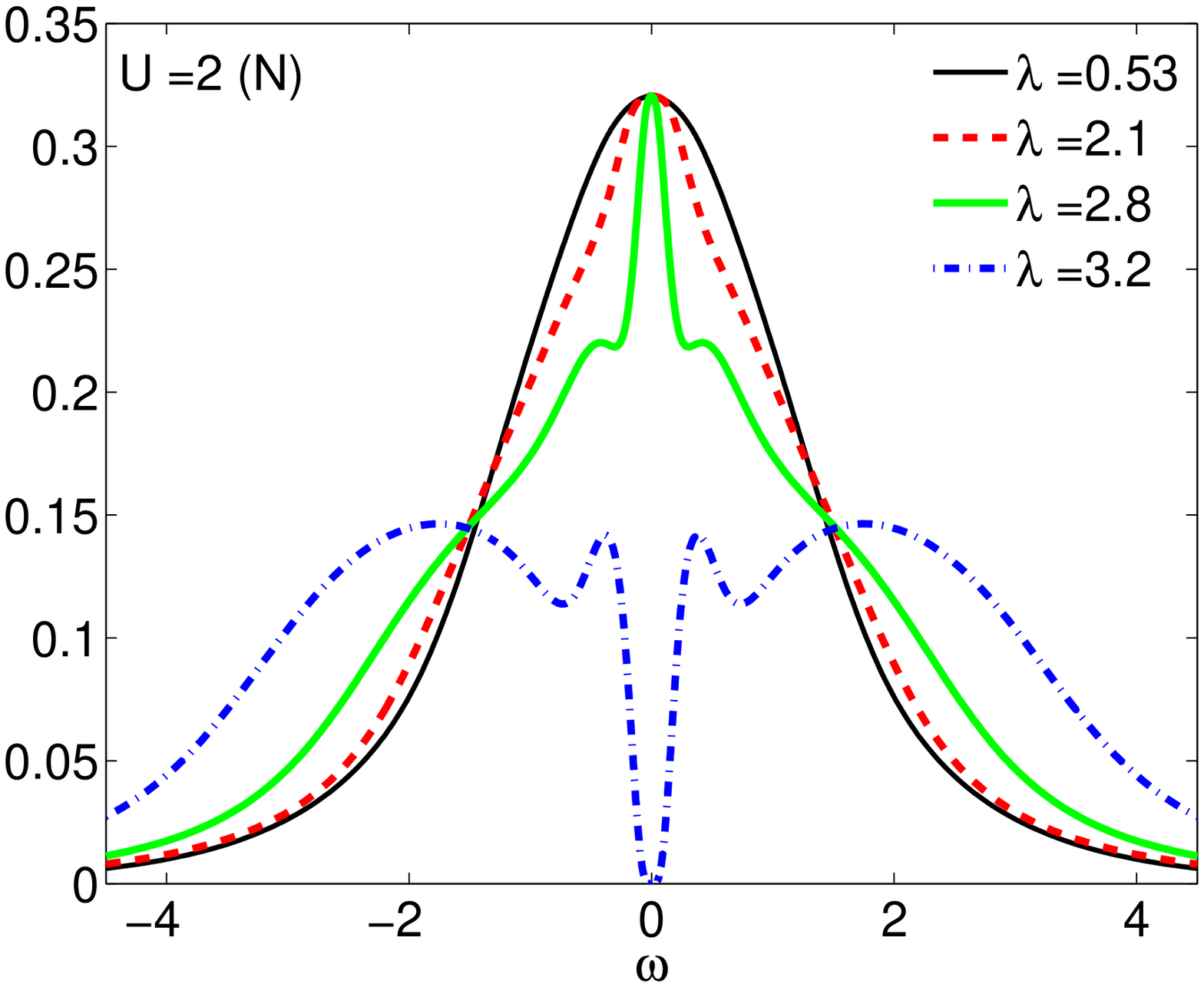}
\includegraphics[width=0.45\textwidth]{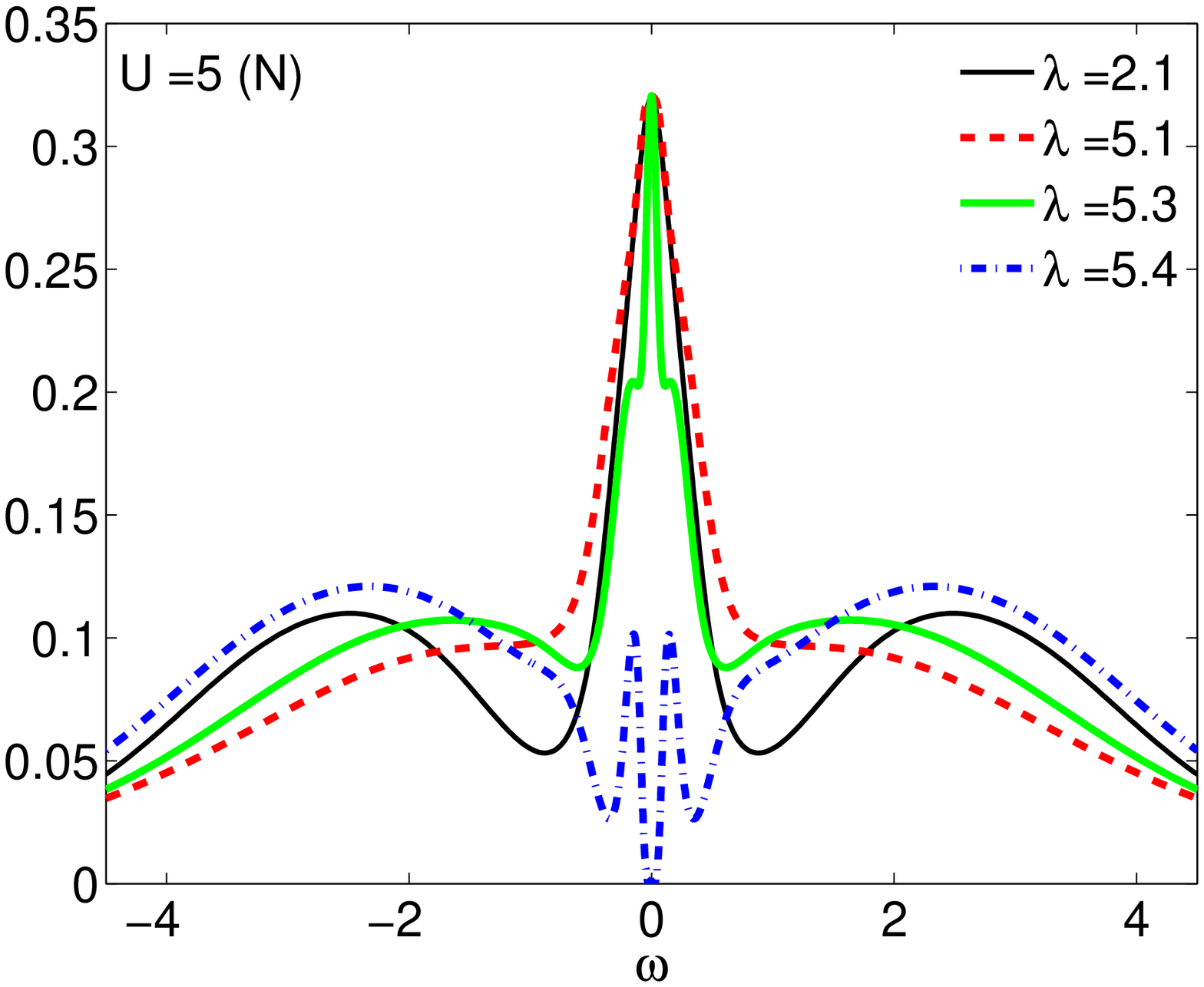}
\caption{(Color online) The local electronic spectral functions $D(\omega)$ in
  comparison for $U=2$ (left panel) and $U=5$ (right panel) for various $\lambda$ and
  $\omega_0=0.6$.}       
\label{elspec_varlambdaU25}
\end{figure}

\noindent
In the left panel we can see for the case of $U=2$ how the
probability distribution becomes broader as $\lambda$ is increased. As 
we do not allow for symmetry breaking here the system changes between zero
occupation and double occupation with the associated oscillator fluctuations
to minimise the energy. The situation is similar to the impurity case shown in
Fig. \ref{rhox1}, and a two peak form develops in a similar way. However in
this case when the two peak structure develops a gap also appears in the
electron density of states, $D(\omega)$, signaling the transition to a
insulating bipolaronic state. The correlation can be seen in the 
corresponding results for $D(\omega)$  shown in Fig.
\ref{elspec_varlambdaU25} over the transition regime.
\noindent

This is in contrast to the impurity density of
states shown in Fig. \ref{rhof} where there is a shift of spectral weight  from the
region near $\omega=0$ to higher and lower values of $\omega$, but a narrow central peak
at $\omega=0$ remains. The narrow peak reflects the fact that there is no
broken symmetry in the impurity case, and there are  fluctuations between the two 
potential wells that restore the symmetry. In the right panel the
 results are shown for $\rho(x)$ across the transition are shown for $U=5$.
There is a similar trend leading to a two peak structure in the bipolaronic
phase, though at larger values of $\lambda$ due to the larger value of $U$.
However, in this case  there is an intermediate regime
where $\rho(x)$ has three maxima, which is very close to the metal-bipolaron
transition, and the change is more marked, occurring  over smaller range of
$\lambda$.\par

\begin{figure}[!htbp]
\centering
\includegraphics[width=0.45\textwidth]{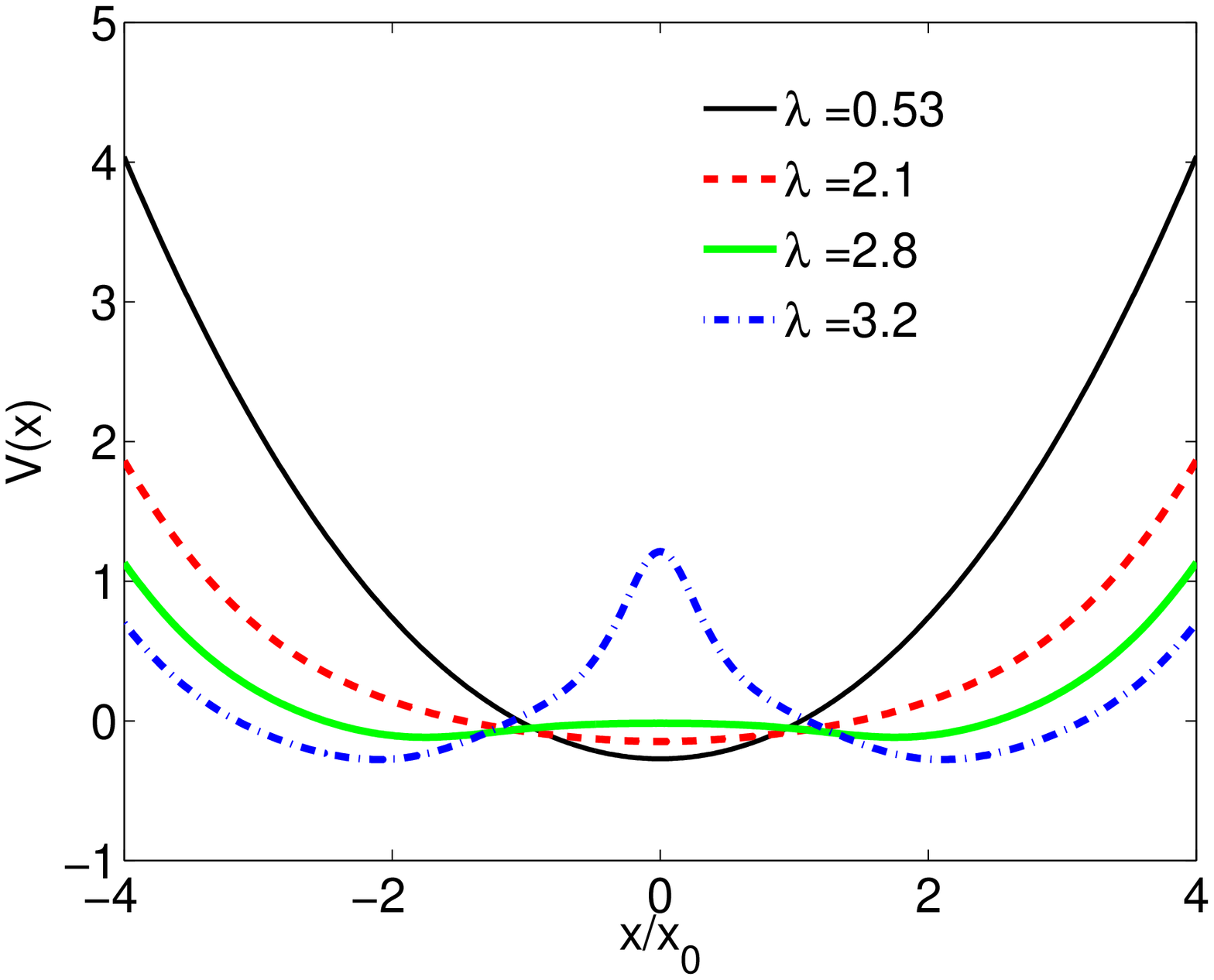}
\includegraphics[width=0.45\textwidth]{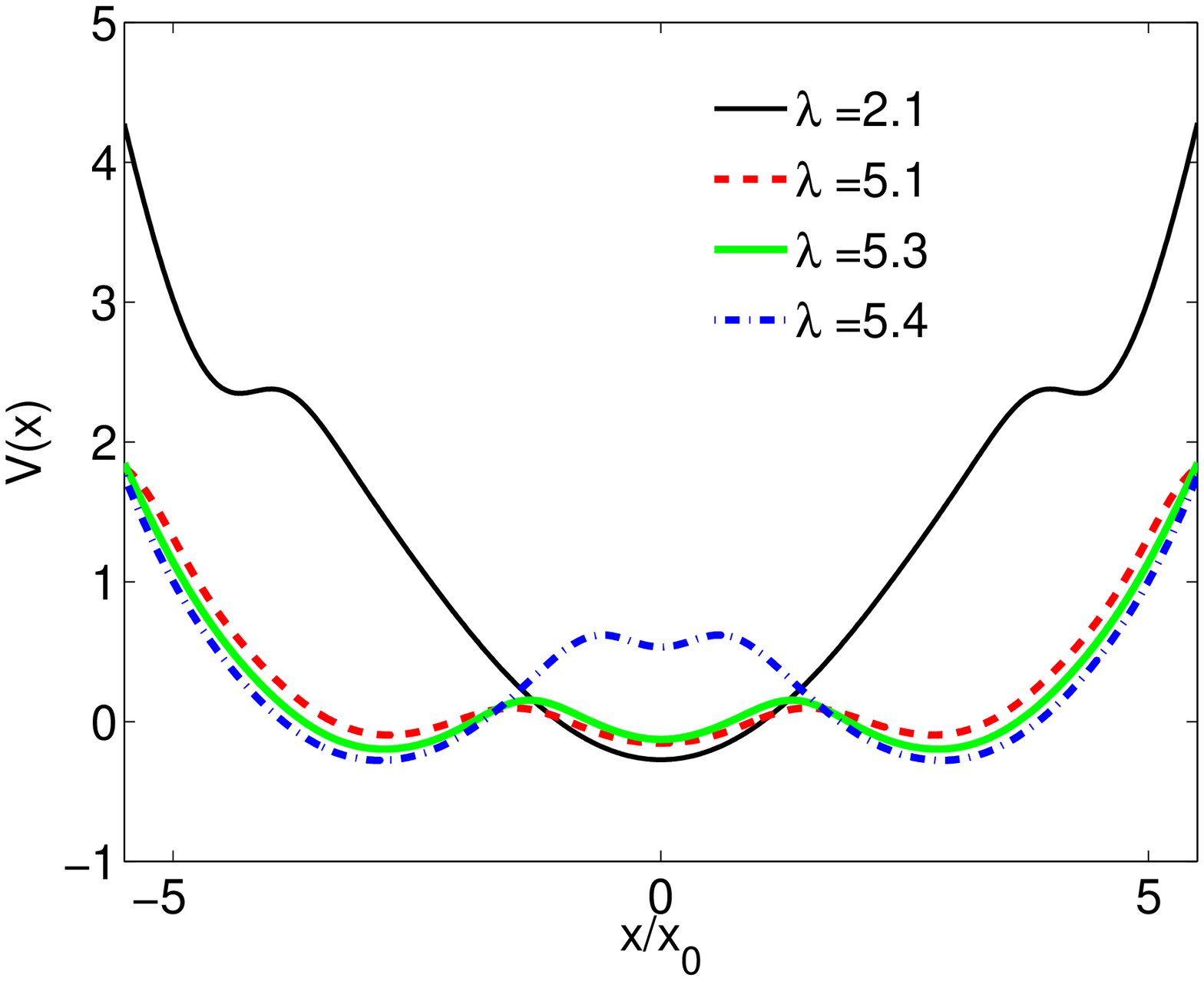}
\caption{(Color online) The local effective potential $V(x)$ for $U=2$
  (left part) for the N state for  $U=5$
  (right part) for various values of $\lambda$ for $\omega_0=0.6$.}       
\label{Vx_xdepvarlambdaN}
\end{figure}

Similar as in the impurity case, Eq. (\ref{pot}), we can compute the effective
potential  $V_{\rm eff}(x)$, which is formed locally in the lattice model due to the
electron-phonon coupling, from the probability distribution. This is shown in
Fig. \ref{Vx_xdepvarlambdaN} for the same values as before, $U=2$ (left
panel) and $U=5$ (right panel). 

\noindent
In the case $U=2$ on increasing $\lambda$ one sees that the potential becomes
shallow and eventually develops two minima at finite $\pm x_m$. Note that
although there are two minima already for $\lambda=2.8$, fluctuations are
sufficient to keep the system in a metallic state, as can be seen also from the
spectral function in Fig. \ref{elspec_varlambdaU25}. When the minima are
deeper, as for $\lambda=3.2$, the system is in the BP insulating state. The
transition is continuous.

In the case $U=5$ the overall trend is similar, but larger values of
$\lambda$ are required to induce the transition. Close to the transition we
can find a structure of 3 local minima, where the one at $x=0$ is lifted upon
increasing $\lambda$. This is characteristic for a discontinuous transition
which is expected to occur for larger values of $U$ as discussed by Koller et
al. \cite{KMH04}.

\begin{figure}[!htbp]
\centering
\includegraphics[width=0.45\textwidth]{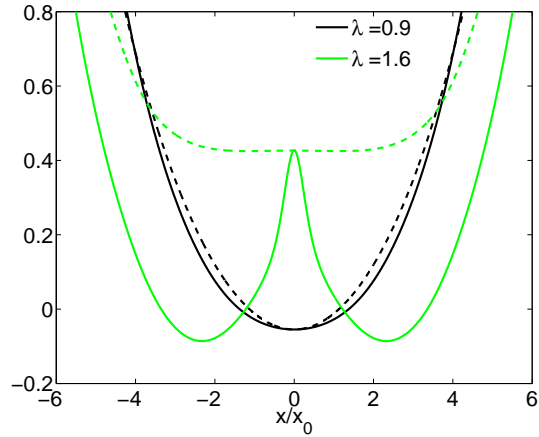}
\caption{(Color online) Comparison of $V(x)$ for DMFT-NRG calculation (full
  line) with the semi-classical approximation (dashed line) for
  $\omega_0=0.2$. The values of $V_{\rm s-cl}(x)$ have been offset to make the
  comparison clearer.}
\label{Vx_complatt}
\end{figure}

If we restrict to the pure Holstein model ($U=0$), then it is also of interest
here to study the quality of the semi-classical approximation.
Similar as in the impurity case the potential can be calculated and one finds
\begin{eqnarray}
V_{\rm
  s-cl}(x)&=&-\frac23 D_0(\bar\mu(x))^3-\bar\mu(x)^2D_0(\bar\mu(x)) \label{potscllatt}\\
&& -\frac{2\bar\mu(x)}{\pi}\arcsin\Big(\frac{\bar\mu(x)}{D}\Big)+\frac12\omega_0^2x^2
-\mu, \nonumber
\end{eqnarray}
where $\bar\mu(x)=\mu-\sqrt{2\omega_0}gx$. The condition $\frac{\partial
  V_{\rm s-cl}(x)}{\partial x}=0$ gives the mean field solutions and one can
infer that at half filling, $\mu=0$, the potential has two minima if
$\lambda>\lambda_c^{\rm mf}=\pi D/4$. The value for this to occur in the DMFT
with local quantum fluctuations is larger and also depends on $\omega_0$. For
$D=2$ one has, e.g., for $\omega_0=0.2$ the value $\lambda_c\simeq 1.75$ and
for $\omega_0=0.6$ the value $\lambda_c\simeq 2$. 
For the first case, $\omega_0=0.2$, we give a comparison of the effective
potential obtained in DMFT calculations and the semiclassical approximation
(\ref{potscllatt}) in Fig. \ref{Vx_complatt} for two values of $\lambda$.

\noindent
For the smaller value of $\lambda$ with the minimum at $x=0$ one finds a quite
good agreement between the calculations. However, closer to the transition the
results vary significantly. $\lambda=1.6$ is still a metallic
solution with a narrow quasiparticle peak in DMFT. As $\lambda>\lambda_c^{\rm
  mf}$ the semiclassical approximation
possesses two shallow minima for this case which can barely be resolved on the
plot. The  positions (1.16 semi-classical, 2.32 DMFT-NRG)  differ
significantly from the DMFT result, where fluctuations keep the state
metallic. Better agreement between for the position of the minima in the
semiclassical approximation and DMFT can again found in the insulating (BP) phase.   

\subsection{Model with long range order}
\begin{figure}[!htbp]
\centering
\includegraphics[width=0.45\textwidth]{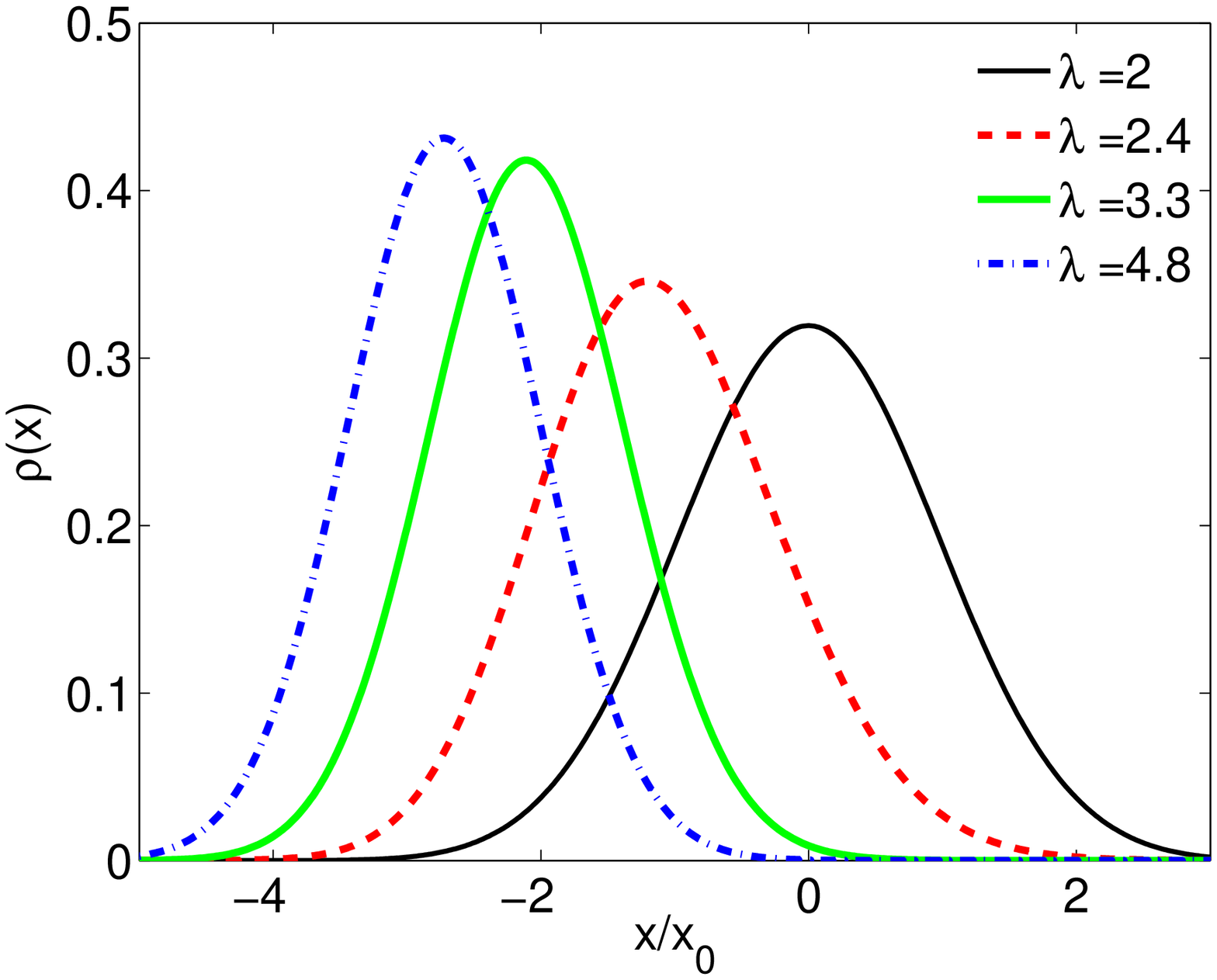}
\includegraphics[width=0.45\textwidth]{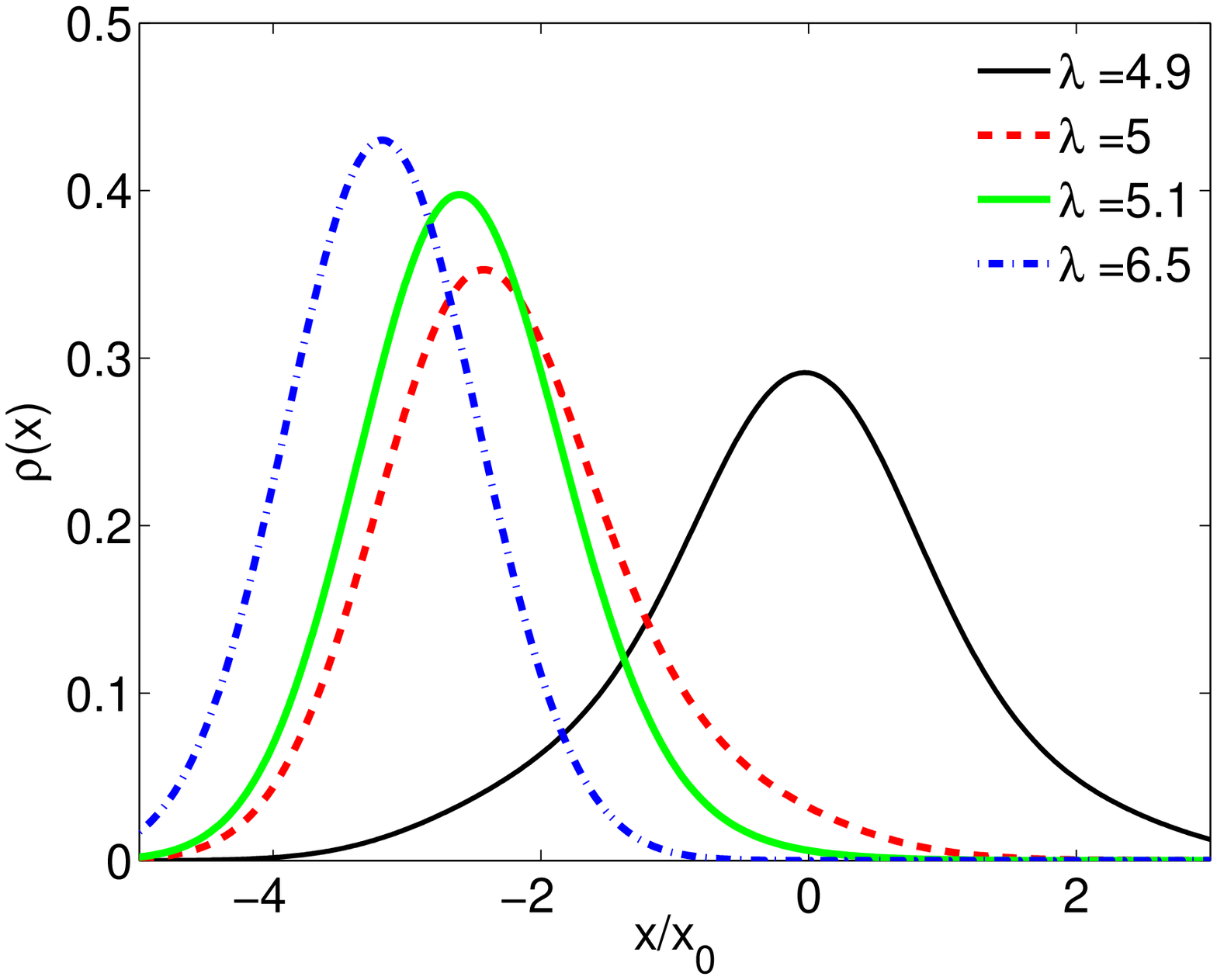}
\caption{(Color online) The local probability distribution $\rho(x)$ for $U=2$
  (left part) for the CO state for  $U=5$
  (right part) for various values of $\lambda$ for $\omega_0=0.6$.}       
\label{rhox_xdepvarlambdaCO}
\end{figure}

To consider charge ordered states, the lattice is divided into two 
sublattices denoted by $A$ and $B$. Charge order develops when the occupation values
$\langle \hat n_{i, A}\rangle\ne \langle \hat n_{i, B}\rangle$,  and their
difference divided by 2 can be taken as the order parameter $\Phi_{\rm co}$. 
As $\langle \hat n_{i, A} +\hat n_{i, B}\rangle=2$ at half filling,  $\Phi_{\rm co}$
can be taken as $1/2(\langle (\hat n_{i, A}\rangle -1)$. When charge order
occurs $\rho(x)$ shifts position to a displaced state appropriate to the local
charge to minimise the energy. This can be seen in Fig.
\ref{rhox_xdepvarlambdaCO} (left panel) where we show $\rho(x)$ for $U=2$ and values of
$\lambda$ as charge order develops for $\lambda>U$. 

\noindent
In this case
$\rho(x)$ has a single peak which shifts and narrows slightly as $\lambda$ is increased.
The same trend can be seen in the right panel of Fig.
\ref{rhox_xdepvarlambdaCO} for the case $U=5$, though the shifting and
narrowing occurs more rapidly as $\lambda$ is increased. The shifting of a
single peak with increasing $\lambda$ is similar to the asymmetric impurity
case shown in Fig. \ref{asymrho}, though the narrowing is an extra feature.
\par
The exact relation in equation (\ref{x}) we derived earlier between the
average displacement and the expectation value for occupation of the impurity
site also holds for the lattice model. If we let $\langle \hat x\rangle$
denote the value of  
$\langle \hat x_{i,A}\rangle$, then  we have from equation (\ref{x})
\begin{equation}
  \Phi_{\rm co}= -{\omega_0\langle \hat x\rangle\over 2\sqrt{\lambda}},
\label{idxphico}
\end{equation}
so that the order parameter is directly proportional to $\langle \hat x\rangle$.
 Again we can test this relation by calculating $\langle \hat x\rangle$ from the 
average over $\rho(x)$ and use the NRG results for  $\Phi_{\rm co}$,
and find that it is satisfied very precisely.

\begin{figure}[!htbp]
\centering
\includegraphics[width=0.45\textwidth]{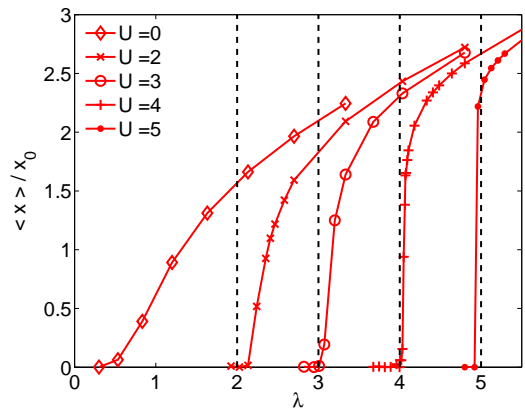}
\caption{(Color online) The expectation values $\expval{x}{}$ for various
  values of $U$ as a function of $\lambda$ for $\omega_0=0.6$ in the CO state.}       
\label{xexp_lamdepvarU}
\end{figure}

In Fig. \ref{xexp_lamdepvarU} we plot $\langle \hat x\rangle/x_0$
for various values of $U$ as a function of $\lambda$, which from
equation (\ref{idxphico}) is proportional to the order parameter $\Phi_{\rm
 co}$. 

\noindent
In the normal or antiferromagnetic state $\langle \hat x\rangle=0$, and
 the onset of the charge order can be seen clearly to occur for $\lambda\sim
 U$. The transition increases sharply with increasing $U$, such that for $U=5$
 it is discontinuous. There is a similar trend in the impurity case shown in
 Fig. \ref{sym}, but it is more marked in the  lattice case.

\begin{figure}[!htbp]
\centering
\includegraphics[width=0.45\textwidth]{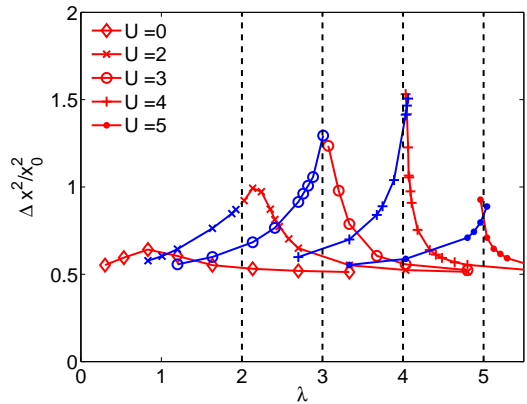}
\caption{(Color online) The expectation values $\Delta x^2$ for various
  values of $U$ as a function of $\lambda$ for $\omega_0=0.6$ in the AFM state
  for $\lambda<U$ and in the CO state for $\lambda>U$.}       
\label{x2mx2exp_lamdepvarU}
\end{figure}

The mean square deviation $(\Delta x)^2$ for the lattice coordinate, which
can be deduced from the appropriate averages over $\rho(x)$, is a measure
of the fluctuations of the order parameter. In Fig. \ref{x2mx2exp_lamdepvarU}
$(\Delta x)^2/x_0^2$  is plotted for the same set of parameters as for
$\langle \hat x\rangle$. 

\noindent
This quantity is finite in the antiferromagnetic phase for $\lambda <U$, and
increases significantly as the transition is approached. The fluctuations can
be seen to become much greater in the region of the transition for the intermediate
values of $U$, most prominently near the point $U=\lambda=4$, where the
transition changes from second to first order. This is a reflection of the
fact that $\rho(x)$ broadens at the transition  and 
then narrows as $\lambda$ is increased further. Once the 
charge order has been well established the fluctuations then fall off rapidly
to give $(\Delta x)^2/x_0^2=1/2$. Though the values of $(\Delta x)^2$ away
from the transition behave like the impurity case shown in Fig. \ref{asym1},
there is a very marked difference in the critical region especially for  large 
$U$. An analysis of the effective potential is also possible for the CO case,
but will be omitted here. For $U=0$ one can similarly derive a semiclassical potential.
A detailed DMFT study of CO order in the Holstein model in the adiabatic limit, where also
the probability distribution and the effective potential are discussed, has
been given by Ciuchi et al. \cite{CP99}.
\par

\section {Conclusions}
We have shown how the reduced density matrix obtained from NRG calculations
can provide physically relevant information about the local fluctuations.
We have illustrated this  by calculating the probability density function $\rho(x)$
for the spatial coordinate $x$ of the local oscillator in the
impurity Anderson-Holstein and  lattice Hubbard-Holstein models. This has
enabled us to address a number of interesting questions.
We have been able to see how the features in $\rho(x)$ correlate
with the features seen in the spectral density of the electrons as the
interaction
strength is increased.
 We have  also deduced an
effective potential $V_{\rm eff}(x)$, such that the wavefunction $|\psi(x)|^2$
of the Schr\"odinger equation corresponds to $\rho(x)$. 
This has enabled us to compare this potential with the one obtained
from a semiclassical approximation, where the $x$-coordinate is
treated as a classical variable, which is equivalent to the Born-Oppenheimer
approximation. The results  have provided a guide as to  when
the semiclassical approximation can be expected to give reliable results,
and to clarify its limitations.\par
 We have also been able to compare the fluctuations of  $x$ in the impurity
 case with those in the lattice model in the various  
parameter regimes. For the normal state BP insulator transition we found a
double well potential for weaker coupling and a structure with three local
minima for stronger coupling. The semiclassical approach only gave a good
description for weaker electron-phonon coupling. Allowing for the symmetry
breaking in the lattice model, we have found that $\rho(x)$ broadens 
in the critical region of the antiferromagnetic to charge order phase
transition. The critical fluctuations become particularly marked in the intermediate
$U$ regime near the point where the ground state transition changes from
continuous to discontinuous behaviour. 

From a calculation of the reduced density matrix it is also possible to  learn something
about the local electronic fluctuations. If in equation (\ref{rho}) we
integrate over the oscillator coordinate $x$, but do not carry out the sum
over $q$ and $m_z$, then we have components $\rho(q,m_z)$ of the impurity
reduced density matrix. From these, for example, we can deduce directly the impurity
charge fluctuation, $(\Delta \hat n_f)^2=\langle (\hat n_f -\langle \hat
n_f\rangle)^2\rangle$,
\begin{equation}
(\Delta \hat n_f)^2=4\rho(2,0)+\sum_{m_z=\pm 1/2}\rho(0,m_z)
-\Big(2\rho(2,0)+\sum_{m_z=\pm 1/2}\rho(0,m_z)\Big)^2.
\end{equation}
If the calculation of the reduced density matrix were to be terminated earlier,
say at the neighbouring site to the impurity, local electronic fluctuations 
and near neighbour correlation functions could be deduced in a similar way.

\bigskip
\noindent{\bf Acknowledgement}\par
\bigskip

 We thank Winfried Koller for his work in initiating this
investigation and for his contribution with Dietrich Meyer
 to the development of the NRG programs. 
\bigskip

\bibliography{artikel}
\bibliographystyle{iopsty}

\end{document}